\documentstyle[11pt]{article}
\input epsf
\begin{document}
\title{
{\bf Theoretical analysis of multi-boson algorithm with 
     local and global update of bosonic fields}
}
\author{Artan Bori\c{c}i\\
        {\normalsize\it Swiss Center for Scintific Computing (SCSC)} \\
        {\normalsize\it Swiss Federal Institute
                        of Technology Z\"urich (ETH)}\\
        {\normalsize\it CH-8092 Z\"urich}\\
        {\normalsize\it borici@scsc.ethz.ch}\\
}
%\date{}
\maketitle
 
\begin{abstract}
We estimate theoretically the cost of the multi-boson method in
the non-hermitian approximation.
It is shown that it is proportional to $V(\log V)^2/m^4$.
For a global update of the scalar fields the cost decreases by a factor $m$
with a $\log V$ overhead.
\end{abstract}

There is an increasing interest in lattice QCD community in better algorithms
for dynamical fermions \cite{Lusch,Jans_Liu,Bunk_etal,BJeg1,BJeg2,
Jans_Liu_BJeg,Sext_Weing,Bor_Forcr,Alexa_etal}.
In \cite{Jans_Liu,Jans_Liu_BJeg} the Kramers algorithm is considered.
This is a variant of the Hybrid Monte Carlo (HMC) algorithm \cite{HMC},
where the equations of motion have a stochastic part. However,
the new algorithm proposed recently by L\"uscher \cite{Lusch} has become
attractive for it brings new views in
full QCD simulations.
The way it is implemented makes it suffer
from the critical slowing down, which is mainly
caused by the local heatbath update
of the bosonic fields \cite{Alexa_etal}.

We want in lattice QCD to estimate the determinant of the quark matrix.
For two degenerate quarks it can be written as $det W^{\dag}W$, where
$W$ is the Wilson matrix with
\begin{equation}
W = \gamma_5 W^{\dag} \gamma_5
\end{equation}
Let $P_n(z), z \in {\bf C}$ be an order $n$ polynomial with
roots, $z_k, k=1,\ldots,n$,
\begin{equation}
P(z) \equiv c_n z^n + . . . + c_1 z + c_0
\end{equation}
such that
\begin{equation}
\frac{1}{z} = \lim_{n \rightarrow \infty} P_n(z)
\end{equation}
Let $R_{n+1}(z)$ be the error of the polynomial approximation,
which is defined as
\begin{equation}
R_{n+1}(z) = 1 - zP_n(z)
\end{equation}
L\"uscher's original proposal uses the hermitian quark matrix
$Q = \gamma_5 W$ and a real polynomial with complex conjugate roots.
As we have proposed in \cite{Bor_Forcr}, a non-hermitian approximation
is expected to work better than the hermitian one.

The method consists in the following equalities:
\begin{equation}
det W^{\dag}W = \lim_{n \rightarrow \infty}
\frac{1}{det P_n(W)^{\dag} P_n(W)}
\end{equation}
\begin{equation}
P_n(z) =  c_{n} \; \prod_{k=1}^{n} (z - z_{k})
\end{equation}
\begin{equation}
\frac{1}{det P_n(W)^{\dag} P_n(W)} = \frac{1}{|c_n|^{2V} (2{\pi}i)^{Vn}}
     \int \prod_{k=1}^{n} [d{\phi}_k^{\dag}] [d{\phi}_k]
     e^{-{\phi}_k^{\dag} (W - z_k)^{\dag}
                         (W - z_k) {\phi}_k }
\end{equation}
where $V \in {\bf N}$ is the rank of $W$.
We use as $P_n(z)$ the Chebyshev polynomials defined in the complex
plane \cite{Bor_Forcr} which have certain optimal properties.

As opposed to HMC, this method introduces a local effective quark action
on both gauge and scalar fields $\phi_k, k=1,\ldots,n$. Naturally,
this allows a local Monte Carlo (MC) update of these fields.
The most important question
is which algorithm is cheaper. We answer this question by theoretical
arguments and propose a global heatbath update for the scalar fields.

We analyse the volume ($V$) and quark mass ($m$) dependence of the cost, which
we denote by $C$ (denoting by $V$ both the volume and the rank of the matrix
should not cause any ambiguity: they are proportional).
Clearly, each MC sweep has a cost proportional to the volume
of the lattice, the number of the bosonic fields and the autocorrelation time
$\tau$. We suppose that the gauge and scalar fields are updated locally
by the heatbath algorithm.
The cost of the algorithm will scale like
\begin{equation}
C \sim V n \tau
\end{equation}
Random walk arguments allow us to assume that
\begin{equation}
\tau \sim \max_k \xi_k^2
\end{equation}
where $\xi_k$ is the correlation length of the operator
$(W - z_k)^{\dag}(W - z_k)$. Then it is clear that
\begin{equation}
\xi_k^2 = \min_k ||[(W - z_k)^{\dag}(W - z_k)]^{-1}||_2
\end{equation}
where $||.||_2$ denotes the 2-norm of a matrix.
To this end we need explicitly the roots of the Chebyshev polynomial
which are given by \cite{Bor_Forcr}
\begin{equation}
z_k = d (1 - \cos \frac{2\pi}{n+1}) - \sqrt{d^2 - c^2} \sin \frac{2\pi}{n+1},
~~~k = 1, \ldots ,n
\end{equation}
where $d > 0$ is the center of the spectrum and $c > 0$ is the focal
distance of the ellipse that encloses the spectrum.
In the asymptotic regime, as $n \rightarrow \infty$, the roots approach
a dense set of points, the ellipse that passes through the origin. Clearly
we obtain
\begin{equation}
\xi_k^2 = ||W^{-1}||_2^2
\end{equation}
which is independent of $k$. Assuming that the smallest singular
value of $W$ behaves like
\begin{equation}
||W^{-1}||_2 \sim m
\end{equation}
we estimate the autocorrelation time to scale like
\begin{equation}
\tau \sim \frac{1}{m^2}
\end{equation}
On the other hand the dynamics of the gauge fields is coupled to
that of the bosonic fields. This can be seen if we look at the
step size of one gauge field update. We use a slightly different
argument of \cite{BJeg2} and write
\begin{equation}
\sum_k {\phi}_k^{\dag} (W - z_k)^{\dag} (W - z_k) {\phi}_k =
\sum_k Tr W^{\dag}W \phi_k \phi_k^{\dag} -
2 Re \sum_k z_k Tr W^{\dag} \phi_k \phi_k^{\dag} +
\sum_k |z_k|^2 Tr \phi_k \phi_k^{\dag}
\end{equation}
We see that for Wilson fermions
the bosonic part of the action is quadratic on gauge fields
$U_i, i \in \{\mbox{set of links}\}$ and the variance of the
corresponding distribution is proportional to
$1 / \sum_k Tr \phi_k \phi_k^{\dag}$. This shows that the step
size of a gauge field update is proportional to $1 / n$.
As result, the autocorrelation time will be proportional to $n$.
Then the total cost of the simulation
will scale like
\begin{equation}
C \sim \frac{V n^2}{m^2}, ~~~n \rightarrow \infty
\end{equation}
To see how the number of boson fields scales with the volume and the quark
mass, we consider the error $\delta$ of the approximation:
\begin{equation}
\delta = det W P_n(W) - 1 = det[I - R_{n+1}(W)] - 1
\end{equation}
or in the eigenvalue basis
\begin{equation}
\delta = \prod_{i=1}^V [1 - R_{n+1}(\lambda_i)] - 1
\end{equation}
Let $M$ be an uniform upper bound for $|R_{n+1}(\lambda_i)|$
(i.e. for each $\lambda_i, i=1,\ldots,V$). Then it can be easily shown that
\begin{equation}
|\delta| \leq (1 - M)^V - 1
\end{equation}
so that in the asymptotic regime ($n$ large and $M$ small) we obtain
\begin{equation}
|\delta| \sim V M
\end{equation}
For Chebyshev polynomials and small $m$ we have \cite{Thesis}
\begin{equation}
M \sim e^{-\alpha m n}, ~~~m \rightarrow 0
\end{equation}
where $\alpha$ is an $O(1)$ real constant. 
In this way we obtain
\begin{equation}
|\delta| \sim V e^{-\alpha m n}, ~~~m \rightarrow 0
\end{equation}
Keeping the approximation error
constant this means that $n$ will scale like
\begin{equation}
n \sim \frac{\log V}{m}
\end{equation}
so that the total cost scales like
\begin{equation}
C \sim \frac{V(\log V)^2}{m^4}, ~~~m \rightarrow 0
\end{equation}
The cost of the HMC algorithm scales at best like \cite{Alexa_etal}
\begin{equation}
C_{HMC} \sim \frac{V^{\frac{5}{4}}}{m^{\frac{13}{4}}}, ~~~m \rightarrow 0
\end{equation}
This shows that the L\"uscher's algorithm scales better with the volume
then HMC, whereas the opposite can be said for the scaling with the
quak mass.
The simulations of dynamical fermions for an $SU(2)$ gauge
theory with the multi-boson algorithm and
Kramers algorithm
show that they perform comparably, the latter being a bit faster
\cite{Jans_Liu_BJeg},
a fact that supports our argument.

However, one can try to reduce the autocorrelation time, so that
the algorithm can compare favourably to HMC. This can be achieved by
performing a global heatbath on the scalar fields.

\subsection*{Global update of bosonic fields}

Consider a global heatbath update of the bosonic fields in the form
\begin{equation}
\phi_k = (W - z_k)^{-1} \eta_k, ~~~\eta_k \sim N(0,I), ~~~k=1,\ldots,n
\end{equation}
This step is very costly because the inversion is not necessary well
conditioned. Instead, we propose a well conditioned inversion to take place:
we use as a polynomial preconditioner the Chebyshev polynomials $P_n(z)$
of the multi-bosonic method and write the above global update as
\begin{equation}
\phi_k = (W - z_k)^{-1} P_n(W) P_n(W)^{-1} \eta_k, ~~~k=1,\ldots,n
\end{equation}
where
\begin{equation}
P_n(W)^{-1} = W [I - R_{n+1}(W)]^{-1}
\end{equation}
The factor $(W - z_k)^{-1}$, has its inverse in
one of the factors of $P_n(W)$, so that we do not need
to compute it. We have to invert instead
the better conditioned matrix $I - R_{n+1}(W)$.
This computation has to take place anyway for the exact version of the
multi-boson algorithm proposed in \cite{Bor_Forcr,Peard_Vienna}.
If $k$ is the number of iterative
steps for the above inversion to converge, its cost will scale like
\begin{equation}
C_{inv} \sim k Vn
\end{equation}
It remains to see how $k$ scales with the volume and the quark mass.
The minimum eigenvalue of the matrix $I - R_{n+1}(W)$ is given by
\begin{equation}
\min_{z \in eig(W)}|I - R_{n+1}(z)| = 1 - \max_{z \in eig(W)} |R_{n+1}(z)|
\end{equation}
We had
\begin{equation}
\max_{z \in eig(W)} |R_{n+1}(z)| \sim e^{-\alpha m n}
\end{equation}
so that we obtain
\begin{equation}
\min_{z \in eig(W)}|I - R_{n+1}(z)| \sim nm, ~~~m \rightarrow 0
\end{equation}
and the number of iterations $k$ scales like
\begin{equation}
k \sim \frac{1}{nm}
\end{equation}
As the quark mass is fixed and $n$ grows,
the matrix $I - R_{n+1}(W)$ is well conditioned.
In any case an optimal iterative
solver requires a minimum number of steps to converge that grows like
$\log V$. This can be seen for example by modeling our well conditioned
problem as an inversion of the quark matrix in one dimension
(i.e. the quantum mechanics of a fermion particle). This problem can be solved
by divide and conquer: by even-odd splitting the original lattice we
obtain two decoupled sublattices, which of them can be split similarly in two
sublattices and so on (the first step as we know can be used
in higher dimensions too). Clearly,
the number of steps needed to arrive at one-site sublattices is
$\log_2V$. This way, we finally get
\begin{equation}
k \sim \frac{\log V}{nm} \sim 1
\end{equation}
and the inversion cost will scale like
\begin{equation}
C_{inv} \sim  Vn \sim \frac{V\log V}{m},
~~~m \rightarrow 0, ~~~n \rightarrow \infty
\end{equation}
The total cost scales like
\begin{equation}
C \sim C_{inv} \tau n
\end{equation}
with $\tau \sim n$, so that finally we obtain
\begin{equation}
C \sim  V n^3 \sim \frac{V(\log V)^3}{m^3},
~~~m \rightarrow 0, ~~~n \rightarrow \infty
\end{equation}

The cost of the algorithm with a global heatbath update
of the scalar fields decreases by a factor $m$. The overhead is a factor
$\log V$ in the volume.
Since the arguments are given in the limiting case $m \rightarrow 0$
and $n \rightarrow \infty$, we expect the simulations
to verify the above cost scaling in this limit.
For moderate masses and number of bosonic fields the algorithm can scale
better.

We note that the above scaling analysis dos not take into account
the prefactors in the cost of the inversion that can make the simulations
expensive. We stress the fact that if the inversion techniques become
less costly, the global heatbath algorithm can be a real alternative
to the local one.

\section*{Prospectives and concluding remarks}

The Hybrid Monte Carlo algorithm has been already explored
in recent years. It can be improved further, as iterative solvers
become more efficient and non-local reversible integrators
can be constructed.
The multi-boson algorithm is new and allows itself for further
improvement. It is a more complex algorithm which has more
degrees of freedom for improvement. One issue is the optimality
of the polynomial. The range of applications is also broader.
We have mentioned in \cite{Bor_Forcr} that one flavor QCD becomes
possible with this algorithm. It has a straightforward application
to the staggered fermions. In finite density QCD simulations, the
sign problem is a long standing problem. With the multi-boson
algorithm is possible to approximate the phase of the quark
determinant for small chemical potential \cite{Bor_Forcr2}.

As illustration of improvement we mention here briefly the
adaptive computation of the optimal polynomial. The proposal
is the following:

Perform quenched simulations until equilibrium and compute
Ritz values $r_m \in {\bf C} , m = 1, \ldots, n+1$
of the quark matrix to the desired order $n$.

Use Ritz values to construct the Ritz polynomial according to
\begin{equation}
R_{n+1}(z) = \prod_{m=1}^{n+1} \frac{z - r_m}{ - r_m}
\end{equation}

Compute zeros of the optimal polynomial
\begin{equation}
P_n(z) = \frac{1 - R_{n+1}(z)}{z}
\end{equation}
and use them as input for the multi-boson algorithm and do not
change them during the simulations.

As illustration we computed in the hermitian approximation
Ritz values of $W^{\dag}W$ by Lanczos algorithm for $n = 18$ for one
quenched $8^3 \times 16$ blocked configuration at $\beta = 6$
and $\kappa = .18 (\kappa_c = .205)$. In Fig. 1 we show
in the complex plane Ritz values together with adaptive roots
of $P_n(z)$. We compare them with the roots of the Chebyshev polynomial.
Comparison between Ritz and Chebyshev polynomials is given in Fig. 2.
For $n=180$ we have repeated the computation and the result is shown
in Fig. 3. In both cases the Ritz polynomial performs better than
the Chebyshev one. It is exactly zero at the first Ritz value.

\subsection*{Acknowledgements}
I would like to thank Professor K. Schilling and his staff for
the warm atmosphere that sorrounded me in HLRZ J\"ulich, where this
work was written. I thank also J. Sexton and H. Hoeber for helpful
suggestions on this work. Finally, I thank Ph. de Frocrand and
A. Galli for discussions on the global heatbath algorithm.
The gauge field was taken form large scale quenched simulation of
TARO collaboration.

\begin{figure}
\centerline{\epsfysize = 8 in \epsffile {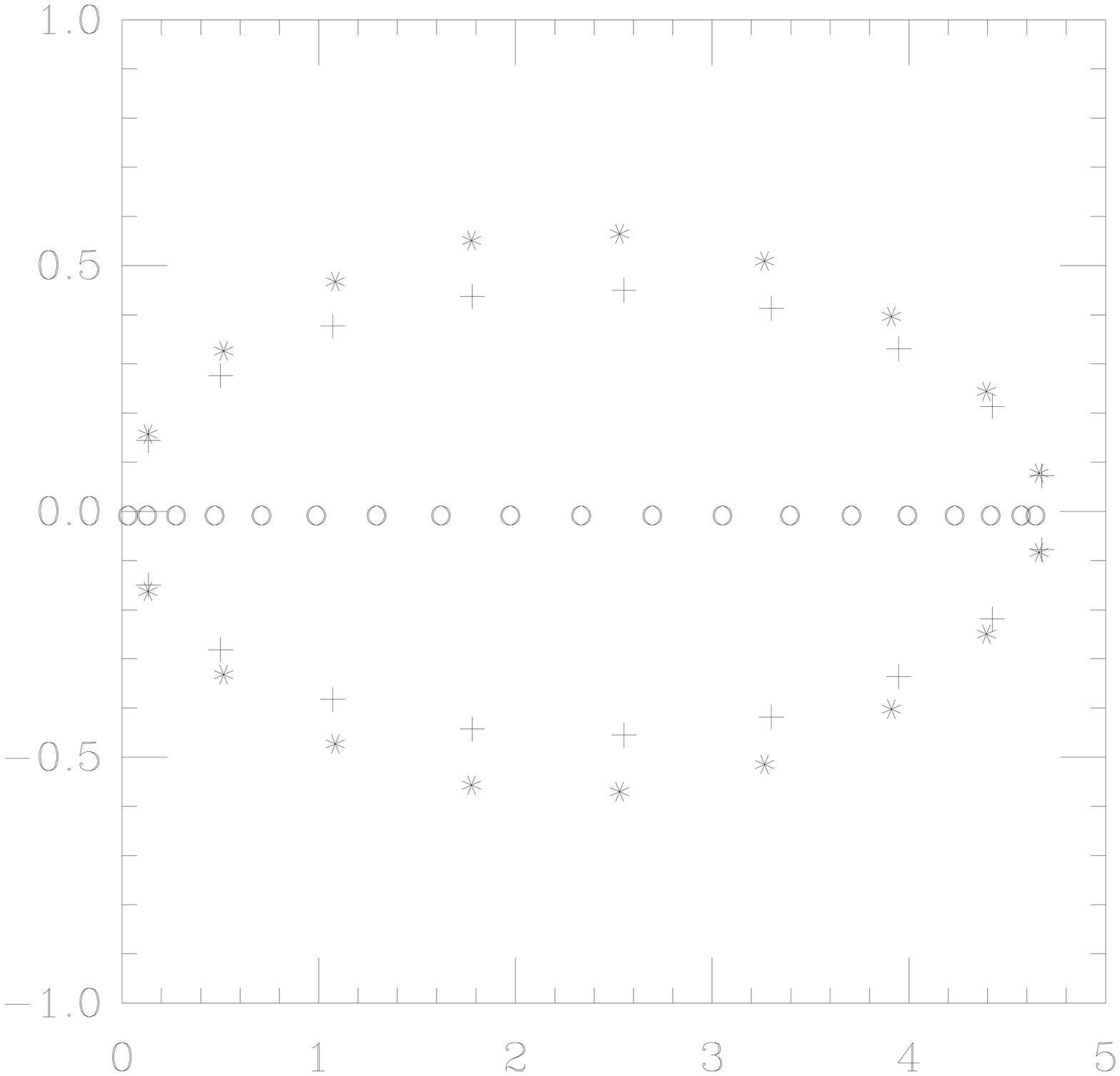}}
\caption{Ritz values (circles) in the complex plane:
$\lambda_{\min} = 0.0442$ and $\lambda_{\max} = 4.6564$.
Stars ($*$) stand for zeros of $P_n(z), n = 18$ computed adaptively,
wheres $+$ for those of Chebyshev polynomials.}
\end{figure}

\begin{figure}
\centerline{\epsfysize = 8 in \epsffile {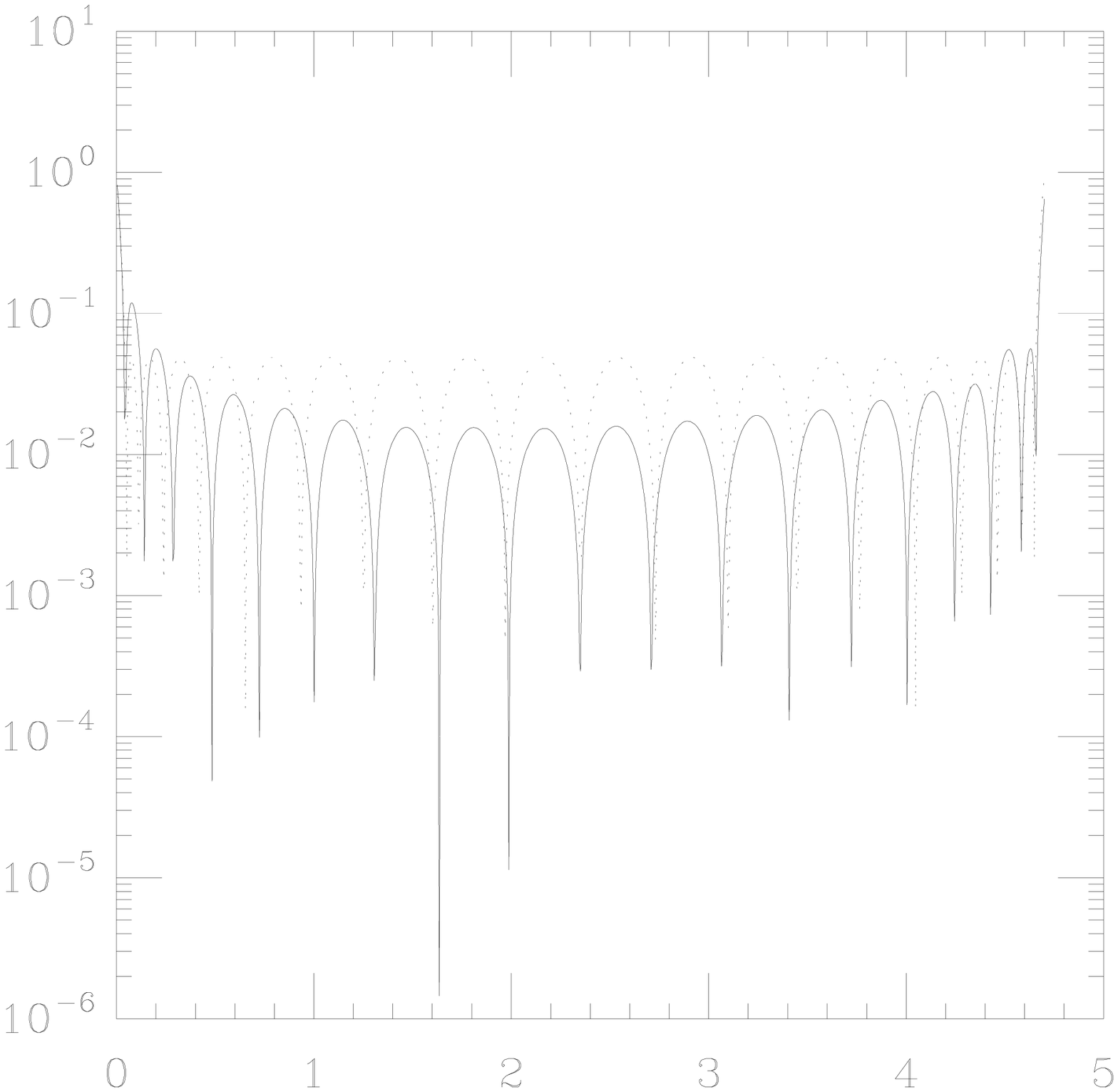}}
\caption{Comparison of Ritz and Chebyshev polynomials in the
hermitian approximation for $n = 18$.}
\end{figure}

\begin{figure}
\centerline{\epsfysize = 8 in \epsffile {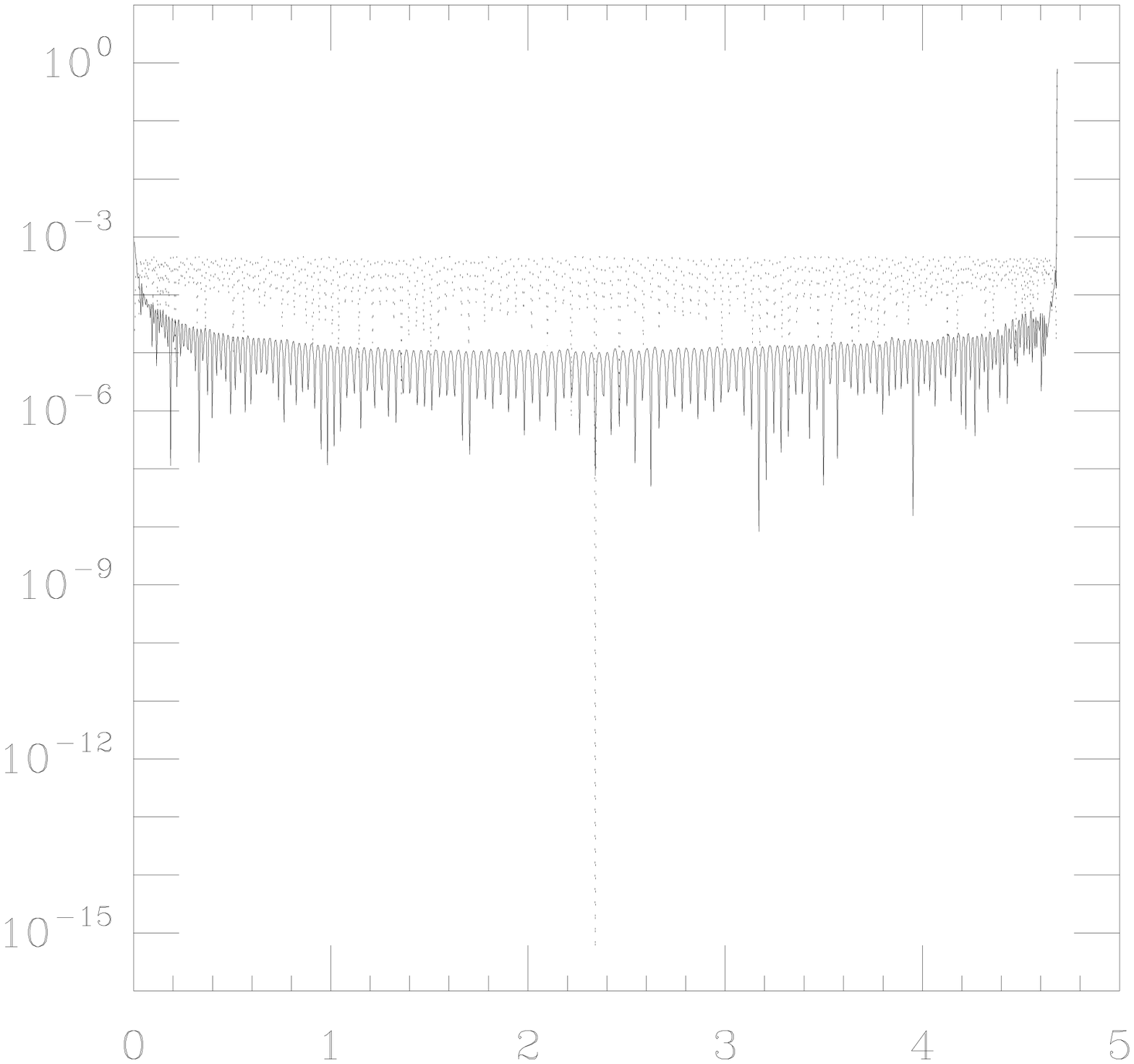}}
\caption{Comparison of Ritz and Chebyshev polynomials in the
hermitian approximation for $n = 180$.}
\end{figure}

\end{document}